
\headline={\ifnum\pageno=1\firstheadline\else
\ifodd\pageno\rightheadline \else\leftheadline\fi\fi}
\def\firstheadline{\hfil} \def\rightheadline{\hfil}
\def\leftheadline{\hfil}
        \footline={\ifnum\pageno=1\firstfootline\else\otherfootline\fi}
\def\firstfootline{\rm\hss\folio\hss}
\def\otherfootline{\hfil}

 1
 1
 1

\font\tenbf=cmbx10
\font\tenrm=cmr10
\font\tenit=cmti10

\font\ninerm=cmr9

\font\eightbf=cmbx8
\font\eightrm=cmr8
\font\eightit=cmti8

\font\sixrm=cmr6 \font\sixi=cmmi6 \font\sixsy=cmsy6 \font\sixbf=cmbx6
\font\sixmsa=msam6 \font\sixmsb=msbm6 \font\sixeufm=eufm6
\font\eightrm=cmr8 \font\eighti=cmmi8 \font\eightsy=cmsy8 \font\eightbf=cmbx8
\font\eighttt=cmtt8 \font\eightit=cmti8 \font\eightsl=cmsl8 \font\eightex=cmex8
\font\eightmsa=msam8 \font\eightmsb=msbm8 \font\eighteufm=eufm8
\font\ninerm=cmr9   \font\ninesy=cmsy9

  \newskip\ttglue

\catcode`@=11
\def\tenpoint{\def\rm{\fam0\tenrm}
\textfont0=\tenrm \scriptfont0=\sevenrm \scriptscriptfont0=\fiverm
\textfont1=\teni \scriptfont1=\seveni \scriptscriptfont1=\fivei
\textfont2=\tensy \scriptfont2=\sevensy \scriptscriptfont2=\fivesy
\textfont3=\tenex \scriptfont3=\tenex \scriptscriptfont3=\tenex
\textfont\itfam=\tenit \def\it{\fam\itfam\tenit}%
\textfont\slfam=\tensl \def\sl{\fam\slfam\tensl}%
\textfont\ttfam=\tentt \def\tt{\fam\ttfam\tentt}%
\textfont\bffam=\tenbf \scriptfont\bffam=\sevenbf
 \scriptscriptfont\bffam=\fivebf \def\bf{\fam\bffam\tenbf}%
\textfont\msafam=\tenmsa\scriptfont\msafam=\sevenmsa
 \scriptscriptfont\msafam=\fivemsa
\textfont\msbfam=\tenmsb \scriptfont\msbfam=\sevenmsb
 \scriptscriptfont\msbfam=\fivemsb
\textfont\eufmfam=\teneufm \scriptfont\eufmfam=\seveneufm
 \scriptscriptfont\eufmfam=\fiveeufm
\tt \ttglue=.5em plus.25em minus.15em
\setbox\strutbox=\hbox{\vrule height8.5pt depth3.5pt width0pt}
\let\sc=\eightrm \let\big=\tenbig \rm}
\def\eightpoint{\def\rm{\fam0\eightrm}
\textfont0=\eightrm \scriptfont0=\sixrm \scriptscriptfont0=\fiverm
\textfont1=\eighti \scriptfont1=\sixi \scriptscriptfont1=\fivei
\textfont2=\eightsy \scriptfont2=\sixsy \scriptscriptfont2=\fivesy
\textfont3=\eightex \scriptfont3=\eightex \scriptscriptfont3=\eightex
\textfont\itfam=\eightit \def\it{\fam\itfam\eightit}%
\textfont\slfam=\eightsl \def\sl{\fam\slfam\eightsl}%
\textfont\ttfam=\eighttt \def\tt{\fam\ttfam\eighttt}%
\textfont\bffam=\eightbf \scriptfont\bffam=\sixbf
 \scriptscriptfont\bffam=\fivebf \def\bf{\fam\bffam\eightbf}%
\textfont\msafam=\eightmsa\scriptfont\msafam=\sixmsa
 \scriptscriptfont\msafam=\fivemsa
\textfont\msbfam=\eightmsb \scriptfont\msbfam=\sixmsb
 \scriptscriptfont\msbfam=\fivemsb
\textfont\eufmfam=\eighteufm \scriptfont\eufmfam=\sixeufm
 \scriptscriptfont\eufmfam=\fiveeufm
\tt \ttglue=.5em plus.25em minus.15em
\setbox\strutbox=\hbox{\vrule height7pt depth2pt width0pt}%
\let\sc=\sixrm \let\big=\eightbig \rm}
\def\tenbig#1{{\hbox{$\left#1\vbox to8.5pt{}\right.\n@space$}}}
\def\eightbig#1{{\hbox{$\textfont0=\ninerm\textfont2=\ninesy
\left#1\vbox to6.5pt{}\right.\n@space$}}}
\catcode`@=12
\font\sevenrm=cmr7  
\parindent=1.2pc
\magnification=\magstep1
\hsize=6.0truein
\vsize=8.6truein
\nopagenumbers

\input tables.tex

\hfuzz=32pt

\catcode`\@=11 

\def\nolabels{\def\wrlabel##1{}\def\eqlabel##1{}\def\reflabel##1{}}
\def\writelabels{\def\wrlabel##1{\leavevmode\vadjust{\rlap{\smash%
{\line{{\escapechar=` \hfill\rlap{\sevenrm\hskip.03in\string##1}}}}}}}%
\def\eqlabel##1{{\escapechar-1\rlap{\sevenrm\hskip.05in\string##1}}}%
\def\thlabel##1{{\escapechar-1\rlap{\sevenrm\hskip.05in\string##1}}}%
\def\reflabel##1{\noexpand\llap{\noexpand\sevenrm\string\string\string##1}}}
\nolabels
\global\newcount\secno \global\secno=0
\global\newcount\meqno \global\meqno=1
\global\newcount\mthno \global\mthno=1
\global\newcount\mexno \global\mexno=1
\global\newcount\mquno \global\mquno=1
\global\newcount\tblno \global\tblno=1
\def\newsec#1{\global\advance\secno by1 
\global\subsecno=0\xdef\secsym{\the\secno.}\global\meqno=1\global\mthno=1
\global\mexno=1\global\mquno=1\global\figno=1\global\tblno=1

\bigbreak\medskip\noindent{\bf\the\secno. #1}\writetoca{{\secsym} {#1}}
\par\nobreak\medskip\nobreak}
\xdef\secsym{}
\global\newcount\subsecno \global\subsecno=0
\def\subsec#1{\global\advance\subsecno by1 \global\subsubsecno=0
\xdef\subsecsym{\the\subsecno.}
\bigbreak\noindent{\it\secsym\the\subsecno. #1}\writetoca{\string\quad
{\secsym\the\subsecno.} {#1}}\par\nobreak\medskip\nobreak}
\xdef\subsecsym{}
\global\newcount\subsubsecno \global\subsubsecno=0
\def\subsubsec#1{\global\advance\subsubsecno by1
\bigbreak\noindent{\it\secsym\the\subsecno.\the\subsubsecno.
                                   #1}\writetoca{\string\quad
{\the\secno.\the\subsecno.\the\subsubsecno.} {#1}}\par\nobreak\medskip\nobreak}
\global\newcount\appsubsecno \global\appsubsecno=0
\def\appsubsec#1{\global\advance\appsubsecno by1 \global\subsubsecno=0
\xdef\appsubsecsym{\the\appsubsecno.}
\bigbreak\noindent{\it\secsym\the\appsubsecno. #1}\writetoca{\string\quad
{\secsym\the\appsubsecno.} {#1}}\par\nobreak\medskip\nobreak}
\xdef\appsubsecsym{}
\def\appendix#1#2{\global\meqno=1\global\mthno=1\global\mexno=1
\global\figno=1\global\tblno=1
\global\subsecno=0\global\subsubsecno=0
\global\appsubsecno=0
\xdef\appname{#1}
\xdef\secsym{\hbox{#1.}}
\bigbreak\bigskip\noindent{\bf Appendix #1. #2}
\writetoca{Appendix {#1.} {#2}}\par\nobreak\medskip\nobreak}
%
%
\def\eqnn#1{\xdef #1{(\secsym\the\meqno)}\writedef{#1\leftbracket#1}%
\global\advance\meqno by1\wrlabel#1}
\def\eqna#1{\xdef #1##1{\hbox{$(\secsym\the\meqno##1)$}}
\writedef{#1\numbersign1\leftbracket#1{\numbersign1}}%
\global\advance\meqno by1\wrlabel{#1$\{\}$}}
\def\eqn#1#2{\xdef #1{(\secsym\the\meqno)}\writedef{#1\leftbracket#1}%
\global\advance\meqno by1$$#2\eqno#1\eqlabel#1$$}
%
%
\def\thm#1{\xdef #1{\secsym\the\mthno}\writedef{#1\leftbracket#1}%
\global\advance\mthno by1\wrlabel#1}
\def\exm#1{\xdef #1{\secsym\the\mexno}\writedef{#1\leftbracket#1}%
\global\advance\mexno by1\wrlabel#1}
%
%
\def\tbl#1{\xdef #1{\secsym\the\tblno}\writedef{#1\leftbracket#1}%
\global\advance\tblno by1\wrlabel#1}
%
\newskip\footskip\footskip14pt plus 1pt minus 1pt 
\def\f@@t{\baselineskip\footskip\bgroup\aftergroup\@foot\let\next}
\setbox\strutbox=\hbox{\vrule height9.5pt depth4.5pt width0pt}
\global\newcount\ftno \global\ftno=0
\def\foot{\global\advance\ftno by1\footnote{$^{\the\ftno}$}}
%
\newwrite\ftfile
\def\footend{\def\foot{\global\advance\ftno by1\chardef\wfile=\ftfile
$^{\the\ftno}$\ifnum\ftno=1\immediate\openout\ftfile=foots.tmp\fi%
\immediate\write\ftfile{\noexpand\smallskip%
\noexpand\item{f\the\ftno:\ }\pctsign}\findarg}%
\def\footatend{\vfill\eject\immediate\closeout\ftfile{\parindent=20pt
\centerline{\bf Footnotes}\nobreak\bigskip\input foots.tmp }}}
\def\footatend{}
%
%
\global\newcount\refno \global\refno=1
\newwrite\rfile
\def\ref{\the\refno\nref}
\def\bref{\nref}
\def\nref#1{\xdef#1{\the\refno}\writedef{#1\leftbracket#1}%
\ifnum\refno=1\immediate\openout\rfile=refs.tmp\fi
\global\advance\refno by1\chardef\wfile=\rfile\immediate
\write\rfile{\noexpand\item{[#1]\ }\reflabel{#1\hskip.31in}\pctsign}\findarg}
\def\findarg#1#{\begingroup\obeylines\newlinechar=`\^^M\pass@rg}
{\obeylines\gdef\pass@rg#1{\writ@line\relax #1^^M\hbox{}^^M}%
\gdef\writ@line#1^^M{\expandafter\toks0\expandafter{\striprel@x #1}%
\edef\next{\the\toks0}\ifx\next\em@rk\let\next=\endgroup\else\ifx\next\empty%
\else\immediate\write\wfile{\the\toks0}\fi\let\next=\writ@line\fi\next\relax}}
\def\striprel@x#1{} \def\em@rk{\hbox{}}
\def\lref{\begingroup\obeylines\lr@f}
\def\lr@f#1#2{\gdef#1{\ref#1{#2}}\endgroup\unskip}

\def\addref#1{\immediate\write\rfile{\noexpand\item{}#1}} 
\def\startrefs#1{\immediate\openout\rfile=refs.tmp\refno=#1}
\def\xref{\expandafter\xr@f}\def\xr@f[#1]{#1}
\def\refs#1{[\r@fs #1{\hbox{}}]}
\def\r@fs#1{\edef\next{#1}\ifx\next\em@rk\def\next{}\else
\ifx\next#1\xref #1\else#1\fi\let\next=\r@fs\fi\next}
%

%
 \newwrite\ffile\global\newcount\figno \global\figno=1
%
%
\def\fig{\the\figno\nfig}
\def\nfig#1{\xdef#1{\secsym\the\figno}%
\writedef{#1\leftbracket \noexpand~\the\figno}%
\ifnum\figno=1\immediate\openout\ffile=figs.tmp\fi\chardef\wfile=\ffile%
\immediate\write\ffile{\noexpand\medskip\noexpand\item{Figure\ \the\figno. }
\reflabel{#1\hskip.55in}\pctsign}\global\advance\figno by1\findarg}
\def\xfig{\expandafter\xf@g}\def\xf@g \penalty\@M\ {}
\def\figs#1{figs.~\f@gs #1{\hbox{}}}
\def\f@gs#1{\edef\next{#1}\ifx\next\em@rk\def\next{}\else
\ifx\next#1\xfig #1\else#1\fi\let\next=\f@gs\fi\next}
%
%
\newwrite\lfile

{\escapechar-1\xdef\pctsign{\string\%}\xdef\leftbracket{\string\{}
\xdef\rightbracket{\string\}}\xdef\numbersign{\string\#}}

\def\writestop{\def\writestoppt{\immediate\write\lfile{\string\pageno%
\the\pageno\string\startrefs\leftbracket\the\refno\rightbracket%
\string\def\string\secsym\leftbracket\secsym\rightbracket%
\string\secno\the\secno\string\meqno\the\meqno}\immediate\closeout\lfile}}
\def\writestoppt{}\def\writedef#1{}

\def\seclab#1{\xdef #1{\the\secno}\writedef{#1\leftbracket#1}\wrlabel{#1=#1}}
\def\applab#1{\xdef #1{\appname}\writedef{#1\leftbracket#1}\wrlabel{#1=#1}}
\def\subseclab#1{\xdef #1{\secsym\the\subsecno}%
\writedef{#1\leftbracket#1}\wrlabel{#1=#1}}
\def\appsubseclab#1{\xdef #1{\secsym\the\appsubsecno}%
\writedef{#1\leftbracket#1}\wrlabel{#1=#1}}
\def\subsubseclab#1{\xdef #1{\secsym\the\subsecno.\the\subsubsecno}%
\writedef{#1\leftbracket#1}\wrlabel{#1=#1}}
\newwrite\tfile \def\writetoca#1{}
\def\leaderfill{\leaders\hbox to 1em{\hss.\hss}\hfill}
\def\writetoc{\immediate\openout\tfile=toc.tmp
   \def\writetoca##1{{\edef\next{\write\tfile{\noindent ##1
   \string\leaderfill {\noexpand\number\pageno} \par}}\next}}}

%
\catcode`\@=12 
%
%
%
%
%
\def\dbend{{\manual\char127}}
\def\d@nger{\medbreak\begingroup\clubpenalty=10000
    \def\par{\endgraf\endgroup\medbreak} \noindent\hang\hangafter=-2
    \hbox to0pt{\hskip-\hangindent\dbend\hfill}\ninepoint}
\outer\def\danger{\d@nger}

\def\p{\partial}

\def\darr#1{\raise1.5ex\hbox{$\leftrightarrow$}\mkern-16.5mu #1}
\def\half{{\textstyle{1\over2}}} 

%
%
\def\al{\alpha}

\def\de{\delta}  \def\De{\Delta}
\def\ep{\epsilon}

\def\la{\lambda} \def\La{\Lambda}
\def\rh{\rho}
\def\si{\sigma}

  \def\Ph{\Phi}

  \def\Om{\Omega}
%
%

%

%
%
 
\def\cC{{\cal C}} \def\cD{{\cal D}}

\def\cI{{\cal I}} 
\def\cL{{\cal L}}

\def\cP{{\cal P}}
\def\cR{{\cal R}} \def\cS{{\cal S}}

\def\cW{{\cal W}}

\def\mapright#1{\smash{\mathop{\longrightarrow}\limits^{#1}}}

\def\ie{{\it i.e.}}
\def\eg{{\it e.g.}}

\def\vps{^{\vphantom{*}}}

\def\Ep{{\cal E}}

%
%
\def\amsyes{y }

\def\answ{y }

\ifx\answ\amsyes
\input amssym.def


\def\CC{{\Bbb C}}
\def\ZZ{{\Bbb Z}}

\def\fC{{\frak C}} 
\def\fH{{\frak H}}
\def\sln{\frak{sl}_N}   
\def\sltw{\frak{sl}_2}  
\def\slth{\frak{sl}_3}  \def\hslth{\widehat{\frak{sl}_3}}

\def\sosi{\frak{so}_6}
\def\soei{\frak{so}_8}
\def\sotwon{\frak{so}_{2N}}

\def\uone{\frak{u}_1}
\def\fA{\frak{A}} \def\bga{\slth\oplus(\uone)^2}
  
\def\fH{\frak{H}} \def\fM{{\frak M}}
\def\fC{\frak{C}} \def\fI{{\frak I}}
\else
\def\ZZ{{Z\!\!\!Z}}
\def\CC{{I\!\!\!\!C}}

\def\sln{s\ell_n}   
\def\sltw{s\ell_2}  
\def\slth{s\ell_3}  \def\hslth{\widehat{s\ell_3}}

\def\sosi{so_6}
\def\soei{so_8}
\def\sotwon{so_{2N}}

\def\uone{u_1}

 \def\bga{{\bf a}}
\def\cR{{\cal R}}
\fi
\def\cG{{\cR}}

%
%

\def\mapright#1{\smash{\mathop{\longrightarrow}\limits^{#1}}}

%
%

%
%
%

\def\AnM#1{Ann.\ Math.\ {\bf #1}}

\def\CMP#1{Comm.\ Math.\ Phys.\ {\bf #1}}

\def\LMP#1{Lett.\ Math.\ Phys.\ {\bf #1}}
\def\LNM#1{Lect.\ Notes in Math.\ {\bf #1}}

\def\NPB#1{Nucl.\ Phys.\ {\bf B#1}}
\def\PLB#1{Phys.\ Lett.\ {\bf {#1}B}}

\def\PRep#1{Phys.\ Rep.\ {\bf #1}}

%

%
%
\def\SMu{\hbox{\lower 3pt\hbox{ \epsffile{su10.eps}}}}
\def\SMs{\hbox{\lower 3pt\hbox{ \epsffile{ss10.eps}}}}
\def\SMd{\hbox{\lower 3pt\hbox{ \epsffile{sd10.eps}}}}

\def\SMS{\leavevmode\vadjust{\rlap{\smash%
{\line{{\escapechar=` \hfill\rlap{\hskip.3in%
                 \hbox{\lower 2pt\hbox{\epsffile{sd10.eps}}}}}}}}}}
\def\SMH{\leavevmode\vadjust{\rlap{\smash%
{\line{{\escapechar=` \hfill\rlap{\hskip.3in%
                 \hbox{\lower 2pt\hbox{\epsffile{su10.eps}}}}}}}}}}
%
%
\def\LW#1{\lower .5pt \hbox{$\scriptstyle #1$}}
\def\LWr#1{\lower 1.5pt \hbox{$\scriptstyle #1$}}
\def\LWrr#1{\lower 2pt \hbox{$\scriptstyle #1$}}
\def\RSr#1{\raise 1pt \hbox{$\scriptstyle #1$}}
\def\cWth{\cW_3}

\def\cWth{{\cal W}_3}

\def\Intro{1}
\def\Intronot{1.1}
\def\eqlat{(1.1)}
\def\AVir{2}
\def\AVcoh{2.1}
\def\sltcoh{2.1}
\def\vircohdec{(2.1)}
\def\AVbv{2.2}
\def\strthsltw{2.2}
\def\defpi{(2.2)}
\def\bvmorph{2.3}
\def\Wthgen{3}
\def\fullcoh{3.1}
\def\bigdecomp{(3.1)}
\def\fullvsprime{(3.2)}
\def\gencones{3.1}
\def\dualcoh{(3.3)}
\def\grconstraint{(3.4)}
\def\maintheorem{3.2}
\def\exactseq{(3.5)}
\def\generpolof{(3.6)}
\def\gmodofi{3.3}
\def\modeldec{(3.7)}
\def\BValgebra{A}
\def\SSprelim{\hbox {A.}1}
\def\BVleib{(\hbox {A.}1)}
\def\BVnother{(\hbox {A.}2)}
\def\SSdefini{\hbox {A.}2}
\def\eqCc{(\hbox {A.}3)}
\def\Smodules{\hbox {A.}3}
\def\gmAE{(\hbox {A.}4)}
\def\SSexamples{\hbox {A.}4}
\def\bvpolyderivations{\hbox {A.}5}
\def\polasder{(\hbox {A.}5)}
\def\vanishideal{(\hbox {A.}6)}
\def\realcond{(\hbox {A.}7)}
\def\generat{\hbox {A.}1}
\def\conone{(\hbox {A.}8)}
\def\contwo{(\hbox {A.}9)}
\def\conthr{(\hbox {A.}10)}
\def\confour{(\hbox {A.}11)}
\def\confive{(\hbox {A.}12)}


\bref\BMPbv{
P.~Bouwknegt, J.~McCarthy and K.~Pilch, {\it The $\cWth$ algebra:
modules, semi-infinite cohomology and BV-algebras},
({\tt hep-th/9509119}).}

\bref\BS{
P.~Bouwknegt and K.~Schoutens, \PRep{223} (1993) 183
({\tt hep-th/9210010}).}

\bref\BSb{
P.~Bouwknegt and K.~Schoutens, {\it $\cW$-symmetry},
Adv.\ Series in Math.\ Phys.\ {\bf 22}, (World Scientific,
Singapore, 1995).}

\bref\BLNW{
M.~Bershadsky, W.~Lerche, D.~Nemeschansky and N.P.~Warner,
\PLB{292} (1992) 35 ({\tt hep-th/9207067}).}

\bref\Wi{
E.~Witten, \NPB{373} (1992) 187 ({\tt hep-th/9108004}).}

\bref\WiZw{
E.~Witten and B.~Zwiebach, \NPB{377} (1992) 55 ({\tt
hep-th/9201056}).}

\bref\LZbv{
B.H.~Lian and G.J.~Zuckerman, \CMP{154} (1993) 613 ({\tt
hep-th/9211072}).}

\bref\LZvir{
B.H. Lian and G.J. Zuckerman, \PLB{254} (1991) 417;
\PLB{266} (1991) 21;  \CMP  {\bf 145} (1992) 561.}

\bref\BMPvir{
P.~Bouwknegt, J.~McCarthy and K.~Pilch, \CMP{145} (1992) 541. }

\bref\WY{
Y.-S.~Wu and C.-J.~Zhu, \NPB{404} (1993) 245 ({\tt hep-th/9209011}).}

\bref\BMPa{
P.~Bouwknegt, J.~McCarthy and K.~Pilch, \LMP{29} (1993) 91
({\tt hep-th/9302086}).}

\bref\BMPb{
P.~Bouwknegt, J.~McCarthy and K.~Pilch, in {\it Perspectives in Mathematical
Physics}, Vol.\ III, {\it eds.} R.~Penner and S.T.~Yau,
(International Press, Boston, 1994)
({\tt hep-th/9303164}).}

\bref\BMPc{
P.~Bouwknegt, J.~McCarthy and K.~Pilch, in the proceedings of the workshop
{\it Quantum Field Theory and String Theory}, Carg\`ese 1993,
{\it eds.\ } L.~Baulieu, V.~Dotsenko, V.~Kazakov and P.~Windey,
NATO ASI Series B: Physics Vol. 328,
(Plenum Press, New York, 1994) ({\tt hep-th/9311137}).}

\bref\BGG{
I.N.~Bernstein, I.M.~Gel'fand and S.I.~Gel'fand,  in
{\it Lie groups and their representations}, Proc.\ Summer School in
Group Representations, Bolyai Janos Math.\ Soc., Budapest 1971,
(Halsted, New York, 1975).}

\bref\Ko{
J.-L.~Koszul, Ast\'erisque (hors s\'erie) (1985) 257.}

\bref\MS{
M.~Penkava and A.~Schwarz, in {\it Perspectives in Mathematical
Physics}, Vol.\ III, {\it eds.} R.~Penner and S.T.~Yau,
(International Press, Boston, 1994) ({\tt hep-th/9212072}).}

\bref\Ak{
F.~Akman, {\it On some generalizations of Batalin-Vilkovisky
algebras}, {\tt (q-alg/9506027)}.}

\bref\Kr{
I.S.~Krasil'shchik, \LNM{1334} (1988) 79.}

\bref\Ge{
M.~Gerstenhaber, \AnM{78} (1962) 267; \AnM{79} (1964) 59.}

\bref\BV{
I.~Batalin and G.~Vilkovisky, Phys.\ Rev.\ {\bf D28} (1983) 2567.}

\bref\Schw{
A.~Schwarz, \CMP{155} (1993) 249, ({\tt hep-th/9205088}).}

\bref\Sch{
J.A.~Schouten, Proc. Ser. {\bf A 43} (1940) 449.}

\bref\Ni{
A.~Nijenhuis, Indag. Math. {\bf 17} (1955) 390.}

\bref\BdFl{
L.C.~Biedenharn and D.E.~Flath,  \CMP{93} (1984) 93.}

%
%
\pageno=0
\line{}
\vskip1cm
\centerline{\bf OPERATOR ALGEBRA OF THE $4D$ $\cWth$ STRING}
\vskip1cm

\centerline{Peter BOUWKNEGT$\,^{1}$, Jim McCARTHY$\,^1$ and
Krzysztof PILCH$\,^2$}
\bigskip

\centerline{\sl $^1$ Department of Physics and Mathematical Physics}
\centerline{\sl University of Adelaide, Adelaide, SA~5005, Australia}
\bigskip

\centerline{\sl $^2$ Department of Physics and Astronomy }
\centerline{\sl  University of Southern California}
\centerline{\sl Los Angeles, CA~90089-0484, USA}
\medskip
\vskip1.5cm

\centerline{\bf ABSTRACT}\medskip
{\rightskip=1cm
\leftskip=1cm
\noindent
The noncritical $4D$ $\cW_3$ string is a
model of $\cW_3$ gravity coupled to two free scalar
fields.  In this paper we discuss its BRST quantization in direct
analogy with that of the $2D$ (Virasoro) string.  The physical
operators form a BV-algebra. We model this BV-algebra on
that of the polyderivations of a commutative ring on six variables
with a quadratic constraint, or, equivalently, on the BV-algebra of
(polynomial) polyvector fields on the base affine space of
$SL(3,\CC)$.}

\vskip2cm
\centerline{To appear in the proceedings of}
\centerline{``STRINGS '95: Future Perspectives in String Theory,''
USC, March 13--18, 1995}

\vfil
\line{USC-95/24 \hfil}
\line{ADP-95-47/M39 \hfil}
\line{{{\tt hep-th/9509121}}\hfil September 1995}

\eject

\footline{\hss \tenrm -- \folio\ -- \hss}

%
%
%
\eightpoint
\centerline{\eightbf OPERATOR ALGEBRA OF THE $4D$
$\cWth$ STRING}
\vglue 0.8cm
%
%
%
\centerline{\eightrm Peter BOUWKNEGT and Jim McCARTHY}
\baselineskip=12pt
\centerline{\eightit Department of Physics and Mathematical Physics,
University of Adelaide}
\baselineskip=10pt
\centerline{\eightit Adelaide, SA~5005, Australia}
\centerline{\eightrm E-mail: pbouwkne@physics.adelaide.edu.au,
jmccarth@physics.adelaide.edu.au}
\vglue 0.2cm
\centerline{\eightrm and}
\vglue 0.2cm
\centerline{\eightrm Krzysztof PILCH}
\baselineskip=12pt
\centerline{\eightit Department of Physics and Astronomy, U.S.C.}
\baselineskip=10pt
\centerline{\eightit Los Angeles, CA~90089-0484, U.S.A.}
\centerline{\eightrm E-mail: pilch@physics.usc.edu}

\vglue 0.6cm
%
%
%
\centerline{\eightrm ABSTRACT}
\vglue 0.2cm
{\rightskip=3pc
 \leftskip=3pc
 \eightrm\baselineskip=10pt\noindent
The noncritical $4D$ $\cW_3$ string is a
model of $\cW_3$ gravity coupled to two free scalar
fields.  In this paper we discuss its BRST quantization in direct
analogy with that of the $2D$ (Virasoro) string.  The physical
operators form a BV-algebra. We model this BV-algebra on
that of the polyderivations of a commutative ring on six variables
with a quadratic constraint, or, equivalently, on the BV-algebra of
(polynomial) polyvector fields on the base affine space of
$SL(3,\CC)$.
\vglue 0.6cm}

\tenpoint\baselineskip=13pt

\newsec{Introduction}
\seclab\Intro

 This paper is a brief outline of our work, over the past few years, on the
$\cW_3$ algebra: its representation theory; the corresponding
semi-infinite cohomology for special modules; the
operator algebra of related $\cW_3$ string models and its
BV-algebra interpretation.  A more complete exposition may
be found in [\BMPbv].
\smallskip

 $\cW_3$ gravities provide interesting examples of gauge theories
based on a nonlinear algebra of constraints; namely, the $\cW_3$
algebra extension of the Virasoro algebra (see [\BS,\BSb] and
references therein).  The nonlinearity immediately leads to a couple
of obvious complications: the adjoint action of the Cartan subalgebra
on $\cWth$ is not diagonalizable, and similarly its action on most
interesting $\cWth$ modules will not be diagonalizable; moreover, the
tensor product of two $\cWth$ modules does not, in general, carry the
structure of a $\cWth$ module.  But still, the algebraic structure
allows a definition of certain $\cWth$ gravity models through BRST
quantization even though the associated $\cW$-geometry is not yet well
understood in general.  In fact, working by analogy with ordinary
two-dimensional gravity, there exists a well-motivated BRST
quantization for $\cW_3$ gravity coupled to conformal matter with a
restricted range for the central charge of the matter CFT.  The
corresponding BRST charge was introduced in [\BLNW].
\smallskip

 Here we will review the special case of $c^M=2$.  In the
corresponding string interpretation the matter scalar fields would
embed the world sheet of the string into a two-dimensional space-time.
But, moreover, since this is a non-critical theory there are dynamical
gravitational degrees of freedom -- under the DDK-type ansatz
these are described by a pair of scalar fields of ``wrong
sign'' with a background charge, the so-called Liouville sector.
Thus, in this string language, the model describes a
$(2+2)$-dimensional string in non-trivial background fields.  We call
it the $4D$ $\cW_3$ string.
\smallskip

More specifically, in this paper we present a geometric model for the
operator algebra of the $4D$ $\cWth$ string, following the
corresponding analysis of the $2D$ (Virasoro) string [\Wi--\LZbv].  We
begin with a summary of the main features of the $2D$ string, in
language which easily generalizes to the $\cWth$ case.  We then
summarize that generalization.  An appendix reviews the notions of
BV-algebras and their modules which are required in this study.

\subsec{Notation}
\subseclab\Intronot

 In the following, $P(\sln)$ and $Q(\sln)$ denote the $\sln$ weight
lattice and root lattice, respectively, while $P_+(\sln)$ denotes the
dominant integral weights.  The matter and Liouville sectors will
be distinguished by a superscript $M$ or $L$, respectively.
$L(\sln) \subset (P(\sln)\times P(\sln))$
is the sublattice defined by
\eqn\eqlat{
L(\sln) =\{(\La^M,-i\La^L)\in P(\sln)\times P(\sln)\, |\,\,
\La^M+i\La^L\in Q(\sln)\}\,.
}
$W(\sln)$ is the Weyl group, $\rho$ always
denotes the corresponding principal Weyl vector, $\ell(w)$
is the length of $w \in W(\sln)$, and $w_0$ is the
reflection in the highest root.  Further, $F^{gh}$ is the Fock space
of the ghost system required for the BRST complex of $\cW_N$-gravity;
\ie, the $j=2,\dots,N$ $(bc)$-systems, $(b^{[j]},\, c^{[j]})$.  Throughout
this paper, we will denote the zero modes of the spin-2 (Virasoro)
ghosts by $b_0, c_0$.

\newsec{The BV-algebra of the $2D$ $\cW_2$ string}
\seclab\AVir

 We first consider the BV-algebra of two-dimensional gravity coupled
to $c^M=1$ matter.  The reader should consult
[\LZvir,\BMPvir,\Wi,\WiZw,\WY], and especially [\LZbv] and the talk by
Zuckerman in these proceedings, for additional discussion and detailed
proofs.

\subsec{The cohomology problem}
\subseclab\AVcoh

 A Fock space representation of the Virasoro algebra, $F(\La,\al_0)$,
is parametrized by the momentum, $\La$, of the underlying Fock space
of a single scalar field with background charge $\al_0$.  If we
interpret $\La$ as an $\sltw$ weight then this representation has
highest (Virasoro) weight $h=\half(\La,\La+2\al_0\rh)$, and the central
charge is given by $c=1-6\al^2_0$.
\smallskip

 An important problem in the study of $2D$ Virasoro string is to
compute the BRST cohomology of the tensor product $F(\La^M,0)\otimes
F(\La^L,2i)$ of Fock spaces with $c=1$ and $c=25$, respectively. More
precisely, we will here introduce the VOA, $\fC$, corresponding to
$\bigoplus_{(\La^M,-i\La^L)\in L(\sltw)}F(\La^M,0)\otimes F(\La^L,2i)\otimes
F^{gh}$.
The BRST complex lifts to $\fC$, on which the differential $d$ acts as
the charge of a spin-1 current.  We will denote by $\fH_{\cW_2} \equiv
H(\cW_2,\fC)$ the cohomology of the complex $(\fC,d)$.  For the
Virasoro case it is useful to define a relative cohomology,
$H_{rel}(\cW_2,\fC)$, for the subcomplex $(\fC,d)_{rel}$ of operators
annihilated by $b_0$ and $L_0^{tot} \equiv \{b_0,d\}$.
\smallskip

 The vertex operator realization of $\widehat\sltw$, together with the
Liouville momentum operator, $-ip^L$, give rise to an $\sltw\oplus
\uone$ symmetry on $\fC$ that commutes with the BRST operator and
yields a decomposition of the cohomology, $H(\cW_2,\fC)$, into a
direct sum of finite dimensional irreducible modules.  The complete
description of the cohomology space is given by the following
theorem due to [\LZvir,\BMPvir,\Wi].

\thm\sltcoh
\proclaim Theorem \sltcoh. The cohomology
$H(\cW_2,\fC)$ is isomorphic, as an $\sltw\oplus \uone$ module, to
the direct sum of doublets of irreducible $\sltw\oplus \uone$ modules with
highest weights in a set of disjoint lines
$\{(\La,\La')+(\la,w^{-1}\la)\,|\,\la\in P_+(\sltw)\}$ labeled by
$w\in W(\sltw) = \{1,r\}$;
\ie,
\eqn\vircohdec{H^n_{rel}(\cW_2,\fC)\ \cong\ \bigoplus_{w\in
W(\sltw)}\,\,\bigoplus_{(\La,\La')\in S_w^n}\,\, \bigoplus_{\la\in P_+(\sltw)}
\cL(\La+\la)\otimes \CC_{\La'+w^{-1}\la}\,,}
where the nontrivial sets of cone tips at gh $=n$, $S_w^n$, are:
$$
\eqalign{
n=0:& \quad\quad S_1^0 = \{(0,0)\} \cr
n=1:& \quad\quad S_1^1 = \{(\rho,-\rho)\}\, ,\quad S_r^1 = \{(0,-2\rho)\} \cr
n=2:& \quad\quad S_r^2 = \{(0,-4\rho)\}\, .\cr
}
$$
\medskip

\noindent
{\it Remarks:}
\item{i.} The doublet structure (note ``doublet'' does
{\it not} refer to an $\sltw$ representation!)
follows from $H^\bullet(\cW_2,\fC)\cong
H_{rel}^{\bullet}(\cW_2,\fC)\oplus H_{rel}^{\bullet-1}(\cW_2,\fC)$
\item{ii.} The Fock space cohomology, $H(\cW_2,F(\La^M,0)\otimes
F(\La^L,2i))$, may be determined either directly (see, \eg, [\BMPvir]),
or by decomposing $F(\La^M,0)$ into irreducible modules and then
computing $H_{rel}(\cW_2,L(\La,0)\otimes F(\La^L,2i))$ (see [\LZvir]).

\subsec{The BV-algebra and structure theorems}
\subseclab\AVbv

 The operator cohomology $(\fH_{\cW_2},\,\cdot\,,b_0)$ forms a
BV-algebra [\LZbv], where the dot product is just the normal-ordered
product of operators in cohomology.  The ground ring $\fH _{\cW_2}^0$
is isomorphic to the algebra of polynomial functions on the complex
plane; \ie, $\fH_{\cW_2}^0\cong \cC_2$, where $\cC_2 = \CC[x_1,x_2]$
is the free Abelian algebra on two generators.  We denote the
corresponding generators of the ground ring by $\widehat x_i$,
$i=1,2$.  It is shown in the appendix that a natural BV-algebra
associated with the ring $\cC_2$ is $(\cP(\cC_2),\,\cdot\,,\De_S)$,
where $\cP(\cC_2)$ is the space of polyderivations of $\cC_2$.  One is
led to ask precisely how these two BV-algebras are related, and this
is answered by the following structure theorem.

\thm\strthsltw
\proclaim Theorem \strthsltw\ [\LZbv].
\item{i.} There is a natural map $\pi:\fH^\bullet_{\cW_2}\longrightarrow
\cP^\bullet(\cC_2)$ which is a BV-algebra homomorphism onto the BV-algebra of
polyderivations $(\cP(\cC_2),\,\cdot\,,\De_S)$.
\item{ii.} There exists an embedding $\imath:\cP(\cC_2)\longrightarrow
\fH_{\cW_2}$ that preserves the dot product and satisfies $\pi\circ\imath={\rm
id}$.

\smallskip

\noindent
{\it Remark:} The projection, $\pi$, is defined by induction on the
degree, $n$. For $n=0$, $\pi$ is the isomorphism discussed above, \ie,
$\pi({\widehat x_i})=x_i$, $i=1,2$. It is then extended to $n>0$ using
the condition (for any $a\in\fH_{\cW_2}^n$ and $x\in\cC_2$)
\eqn\defpi{
\pi(a)(x) ~=~ \pi([a,{\widehat x}])\,.
}

 This theorem is a neat summary of the discussions of [\Wi,\WiZw]
on the geometric interpretation of a part of the cohomology.  One may
proceed to a detailed understanding of the full cohomology.
Let us denote $(\fH_{\cW_2})_+\equiv\imath(\cP(\cC_2))$
and $(\fH_{\cW_2})_-\equiv{\rm Ker}\, \pi$, then
$\fH_{\cW_2}\cong(\fH_{\cW_2})_+\oplus (\fH_{\cW_2})_-$.  There is a
unique element in $(\fH_{\cW_2})_+$, $\widehat \Om$ of ghost number two
at the weight $(0,-2\rh)$, for which $b_0\widehat
\Om\not\in\imath(\cP(\cC_2))$.
By studying the action of $\fH_{\cW_2}$ on its BV-ideal
$(\fH_{\cW_2})_-$, Lian and Zuckerman [\LZbv] conclude that
the BV-algebra $(\fH_{\cW_2},\,,\cdot\,,b_0)$ is generated by $1$, the ground
ring generators $\widehat x_i$ and $\widehat \Om$.
The polyderivation $\Om = \pi(\widehat \Om)$ is the unique nontrivial
homology class of the BV-operator $\De_S$.  Thus, the ``gluing'' of
$(\fH_{\cW_2})_+$ and $(\fH_{\cW_2})_-$, accomplished by the BV-operator
$b_0$, is underlined by a simple algebraic principle:
\smallskip

\noindent
The BV-algebra $(\fH_{\cW_2},\,\cdot\,,b_0)$ is the minimal BV-algebra
extension of $(\cP(\cC_2),\,\cdot\,,\De_S)$ in which the cohomology of
the BV-operator $b_0$ is trivial.
\bigskip

 There is also a nice geometric understanding for the BV-module
$(\fH_{\cW_2})_-$.  Introduce the ``dual'' ring module $M_r$ as
in the appendix.  It is shown in [\LZbv] that $(\fH_{\cW_2})_-^1$ is
isomorphic as a ground ring module to $M_{r}$.  In the appendix we see
that there is a natural BV-module of $\cP(\cC_2)$ whose lowest grade
space is $M_r$: namely, the twisted polyderivations $\cP(\cC_2,M_r)$.

\thm\bvmorph
\proclaim Theorem \bvmorph.  There is a natural map
$\pi':(\fH_{\cW_2})^\bullet_-\rightarrow\cP^{\bullet-1}(\cC_2,M_r)$ which
is a BV-morphism of BV-modules.

\medskip

\noindent
{\it Remark:}
The projection, $\pi'$, is again defined by induction on the degree,
using the isomorphism discussed above for $n=1$ and extending to $n>1$
using \defpi.

\newsec{The BV-algebra of the $4D$ $\cW_3$ string}
\seclab\Wthgen

 In this section we will simply list the appropriate changes which
distinguish the corresponding results for the $\cWth$ string, for
which the underlying problem is the computation of the BRST cohomology of
the tensor product $F(\La^M,0)\otimes F(\La^L,2i)$ of Fock spaces with
$c^M=2$ and $c^L=98$.  Many of these differences can be
traced to the nature of the $\cWth$ algebra and its representation
theory which was briefly touched on in the introduction.  For a
complete discussion, the reader should consult [\BMPbv].
\smallskip

 The immediate difference is that the $\sltw$ structure apparent in
the Virasoro case is replaced by $\slth$.  In particular, a Fock space
representation of the $\cWth$ algebra is now labeled by two components
of momentum, which can be identified with an $\slth$ weight.  We again
lift the cohomology problem to the VOA, $\fC$, corresponding to
$\bigoplus_{(\La^M,-i\La^L)\in L(\slth)}F(\La^M,0)\otimes
F(\La^L,2i)\otimes F^{gh}$.  The cohomology of the complex $(\fC,d)$
is now denoted by $\fH_{\cW_3}\equiv H(\cW_3,\fC)$.  The vertex
operator realization of $\hslth$ on the $c=2$ Fock spaces, together
with the two Liouville momenta, provide a realization of $\bga$ on
$\fC$.
\smallskip

 Given that there are now two antighost zero modes, $b_0 = b_0^{[2]}$
and $b_0^{[3]}$, it might seem natural that the cohomology should
display a quartet rather than doublet structure.  This expectation is
in fact realized.  We call the lowest ghost number state in a given
quartet the ``prime'' state, and denote by $H_{\rm pr}(\cW_3,\fC)$ the
space of prime states.  However, the situation is actually rather
subtle -- in particular, it turns out that $b_0^{[3]}$ is not
well-defined on the cohomology.  Thus, although relative cohomology
can be defined it is not obviously useful -- the nondiagonalizability
of the operator $W_0^{tot} \equiv\{b_0^{[3]},d\}$ makes it extremely
difficult to analyze.
As a result, unlike the doublets of the Virasoro case, the quartet
decomposition is at the level of vector spaces only, and there is no
obvious intrinsic characterization of prime states as specific
cohomology classes.
\smallskip

 Finally, the cohomology will lie in cones, rather than
lines, now labeled by the Weyl group $W(\slth)$.   In place of Theorem
\sltcoh\ we therefore have
\thm\fullcoh
\proclaim Theorem \fullcoh.  The cohomology
$H(\cWth,\fC)$ is isomorphic, as an $\bga$ module, to the
direct sum of quartets
of irreducible $\bga$ modules with the highest weights in a set of
disjoint cones $\{ (\La,\La') + (\la,w^{-1}\la)\,|\,\la\in P_+(\slth),\,
(\La,\La') \in \cS_w\}$ labeled by $w\in W(\slth)$; \ie,
\eqn\bigdecomp{
H_{\rm pr}^n(\cW_3,\fC)\ \cong\ \bigoplus_{w\in W(\slth)}\,
  \bigoplus_{(\La,\La')\in
  \cS_w^n} \, \bigoplus_{\la\in P_+} \left( \cL(\La+\la) \otimes
  \CC_{\La' + w^{-1}\la} \right) \,.}
\smallskip
\noindent
where the sets $\cS_w^n$ (tips of the cones) are listed in Table
\gencones, and
\eqn\fullvsprime{
H^n~\cong~ H^n_{pr}\oplus H^{n-1}_{pr} \oplus
H^{n-1}_{pr} \oplus H^{n-2}_{pr}\,.
}

\vfil\eject

\tbl\gencones

\begintable
\quad $n$ \quad | \quad $w$ \quad | \quad $(\La,\La')$ \quad \crthick
$0$ | $1$ |  $(0,0)$  \cr
$1$ | $1$ |  $(\La_2,\La_1-\La_2)$,
               $(\La_1+\La_2,0)$,
             $(\La_1,-\La_1+\La_2)$ \nr
    | $r_1$ | $(0,-2\La_1+\La_2)$ \nr
    | $r_2$ | $(0,\La_1-2\La_2)$ \cr
$2$ | $1$ | $(2\La_2,-\La_2)$,
            $(0,-\La_1-\La_2)$, $(2\La_1,-\La_1)$ \nr
    | $r_1$ | $(\La_1,-2\La_1)$, $(\La_2,-3\La_1+\La_2)$,
              $(0,-4\La_1+2\La_2)$ \nr
    | $r_2$ | $(\La_2,-2\La_2)$, $(\La_1,\La_1-3\La_2)$,
              $(0,2\La_1-4\La_2)$ \nr
    | $r_{12}$ | $(0,-3\La_2)$ \nr
    | $r_{21}$ | $(0,-3\La_1)$ \cr
$3$ | $1$ | $(\La_1+\La_2,-\La_1-\La_2)$ \nr
    | $r_1$ |  $(\La_2,-2\La_1-\La_2)$, $(\La_1,-4\La_1+\La_2)$,
              $(\La_2,-5\La_1+2\La_2)$  \nr
    | $r_2$ | $(\La_1,-\La_1-2\La_2)$, $(\La_2,\La_1-4\La_2)$,
              $(\La_1,2\La_1-5\La_2)$ \nr
    | $r_{12}$ | $(\La_2,-\La_1-3\La_2)$, $(0,\La_1-5\La_2)$,
                 $(\La_2,-5\La_2)$ \nr
    | $r_{21}$ | $(\La_1,-3\La_1-\La_2)$, $(0,-5\La_1+\La_2)$,
                 $(\La_1,-5\La_1)$ \nr
    | $r_3$ | $(0,-2\La_1-2\La_2)$ \cr
$4$ |  $r_1$ | $(0,-4\La_1-\La_2)$ \nr
    | $r_2$ | $(0,-\La_1-4\La_2)$ \nr
    | $r_{12}$ | $(\La_2,-2\La_1-4\La_2)$, $(\La_1,-\La_1-5\La_2)$,
                 $(0,-6\La_2)$ \nr
    | $r_{21}$ | $(\La_1,-4\La_1-2\La_2)$, $(\La_2,-5\La_1-\La_2)$,
                 $(0,-6\La_1)$ \nr
    | $r_3$ |  $(0,-3\La_1-3\La_2)$, $(2\La_1,-4\La_1-3\La_2)$,
                $(2\La_2,-3\La_1-4\La_2)$ \cr
$5$
    | $r_{12}$ | $(0,-2\La_1-5\La_2)$ \nr
    | $r_{21}$ | $(0,-5\La_1-2\La_2)$ \nr
    | $r_3$ |  $(\La_1,-5\La_1-3\La_2)$,
$(\La_1+\La_2,-4\La_1-4\La_2)$,
        $(\La_2,-3\La_1-5\La_2)$ \cr
$6$ | $r_3$ | $(0,-4\La_1-4\La_2)$ \endtable

\bigskip
\centerline{\it Table \gencones.\ The sets $\cS_w^n$}
\vfil\eject

\noindent
{\it Remarks:}
\item{i.} The Fock space cohomology, $H(\cW_3,F(\La^M,0)\otimes
F(\La^L,2i))$, has not been determined directly.  We have computed it
[\BMPa--\BMPc,\BMPbv] for $-i\La^L+2\rh\in P_+(\slth)$ by decomposing
the Fock space $F(\La^M,0)$ into irreducible modules, and then computing
$H(\cW_3,L(\La,0)\, \otimes F(\La^L,2i))$ using (generalized) Verma
module resolutions.  The latter cohomology displays qualitatively new
features, which, in this particular case, can be explained by the more
complicated submodule structure of Verma modules of a higher rank
$\cW$-algebra.
\item{ii.} Let $\fC_w \subset \fC$ denote the subspace with
shifted Liouville momentum $-i\La^L+2\rh\in w^{-1} P_+(\slth)$.  In
the Virasoro case it is enough to know the cohomology for $\fC_1$, the
fundamental Weyl chamber, since the remainder is related by a
nondegenerate pairing (duality).  For the $\cWth$ case there is again such
a pairing,
\eqn\dualcoh{
H^\bullet(\cW_3,F(\La^M,0)\otimes F(\La^L,2i))~\cong~
H^{8-\bullet}(\cW_3,F(\La^M,0)\otimes F(w_0\cdot\La^L,2i))\,,
}
for all $(\La^M,-i\La^L)\in L$, where $w_0\cdot\La^L = w_0(\La^L +
2i\rho)-2i\rho$, but it only relates the cohomology in $\fC_w$ and
$\fC_{w_0w}$ and thus is not sufficient to deduce the whole cohomology
from $H(\cWth,\fC_1)$.  The Fock space cohomology for $-i\La^L+2\rh$
in the other Weyl chambers has been derived from the assumption of a
kind of Weyl group symmetry [\BMPbv].
\smallskip

  We may now study $(\fH_{\cWth},\,\cdot\,,b_0)$ as a BV-algebra.  The
ground ring $\fH_{\cWth}^0$ in this case is the associative Abelian
algebra generated by the identity operator $1$, together with the
$\slth$ triplet and anti-triplet operators $\widehat x_\si$ and
$\widehat x^\si$, $\si=1,2,3$, subject to the constraint
\eqn\grconstraint{
 \widehat x_\si\cdot \widehat x^\si ~=~ 0\,.
}
Thus we see that $\fH_{\cWth}^0$ is isomorphic to the ring $\cG_3$
discussed in the appendix, and this allows us to present a geometric
model for $\fH_{\cWth}$. Define the natural map $\pi$ between
$(\fH_{\cWth},\,\cdot\,,b_0)$ and $(\cP(\cG_3),\,\cdot \,,\De_S)$ using
\defpi.

\thm\maintheorem
\proclaim Theorem \maintheorem\ [\BMPbv].
\item{i.} The map $\pi:\fH_{\cWth}^\bullet\rightarrow\cP^\bullet(\cG_3)$
is a BV-algebra homomorphism.
\item{ii.} $\frak{I} \equiv {\rm Ker\,}\pi$ is a BV-ideal of
$\fH_{\cWth}$.  The exact sequence of BV-algebras
\eqn\exactseq{
\matrix{
0&\mapright{}&\frak{I}&\mapright{}&\fH_{\cWth}&\mapright{\pi}&
\cP(\cG_3)&\mapright{}&0\,.\cr
}
}
splits as a sequence of $\imath(\cP(\cG_3))$ dot modules, where
$\imath:\cP(\cG_3)\rightarrow \fH_{\cWth}$ is a dot algebra homomorphism
such that $\pi\circ\imath={\rm id}$.
\item{iii.} The cohomology of $b_0$ on $\fH_{\cWth}$ is trivial.
\smallskip

 The last statement implies that the BV-algebra of cohomology is an
extension of the BV-algebra of polyderivations in a way somewhat
similar to the Virasoro case.  However, the kernel $\frak{I}$ is now
rather large, containing, in particular, most operators in
$H(\cWth,\fC_w),\, w\neq1$.  Further, unlike in the Virasoro case, the
bracket and product on $\fI$ are nontrivial.  Moreover, as seen above,
the pairing \dualcoh\ no longer determines the remaining structure.  Thus
it is important to obtain the analogue of Theorem \bvmorph.  As
discussed in the appendix, there are now 6 natural ring module
structures $M_w$, labeled by the Weyl group.  For a given $\fC_w$,
$w\in W(\slth)$, the smallest ghost number with nontrivial cohomology
is $\ell(w)$ where precisely one cone appears, $\widehat M_{w}$, which
is isomorphic as a ring module with $M_w$ [\BMPbv].  In fact, there is
a natural map from $\frak{I}$ to twisted polyderivations, which we
denote by $\pi_w$,
\eqn\generpolof{
\pi_w(\Ph)( x_{i_1},\ldots, x_{i_n}) ~=~ \pi_w(
[\,\ldots\, [\,[\,\Ph,\widehat x_{i_1}\,],\ldots\,],\widehat
x_{i_n}\,]) \,,
}
that identifies $\widehat M_w$ and $M_w$, and maps
$\Phi\in\fH_{\cWth}^{\ell(w)+n}$, with $-i\La^L+2\rh$ sufficiently
deep inside $w^{-1}P_+(\slth)$, onto a twisted polyderivation
$\pi_w(\Phi)\in\cP^n(\cG_3,M_w)$
\smallskip

 Thus, sufficiently far from the overlaps of the Weyl chambers of
shifted Liouville momentum, the cohomology can be identified with the
twisted polyderivations of $\cG_3$.  It remains to characterize just
how the different regions are patched together.  Since the kernel
$\frak{I}$ is a BV-ideal of $\fH_{\cWth}$, it is a BV-module and
therefore a G-module.  Precise calculations show that $\fI^n=0$ for
$n<1$, and $\fI^1\cong \widehat M_{r_1}\oplus \widehat M_{r_2}$.  As
before, we may construct the natural map $\pi'\equiv\pi_{r_1}\oplus
\pi_{r_2}\,:\, \fI^n \rightarrow
\cP^{n-1}(\cR_3,M_{r_1}) \oplus \cP^{n-1}(\cR_3,M_{r_2})$,
which is equal to the identity on $\fI^1$ and for $n\geq 2$ is given
by the multiple brackets \generpolof.

\thm\gmodofi
\proclaim Theorem \gmodofi\ [\BMPbv]. The map $\pi'$ is a G-morphism of
G-modules.

\noindent
We conjecture that this holds at the level of BV-modules.
The remainder of the structure of the cohomology is now
determined by duality.  For details, the reader should consult [\BMPbv].
\smallskip

The geometrical aspects of the cohomology are more manifest if we
realize $\cP(\cR_3)$ as the space of regular polyvector fields on
the base affine space of $SL(3,\CC)$.
\smallskip

 Following [\BGG], the base affine space of $SL(3,\CC)$ is defined as
the quotient $A=N_+\backslash SL(3,\CC)$, where $N_+$ is the nilpotent
subroup generated by the positive root generators.  The space of regular
functions on $A$, $\Ep(A)$, consists of those functions in $\Ep(G)$ that
are invariant under\foot{We will distinguish between
left and right actions by labels $L$ and $R$.} $N_+^L$, and carries a
representation of $(\slth)_R\oplus(\uone^2)_L$.
It is an immediate consequence of the Peter-Weyl theorem that
$\Ep(A)$ is a model space for $\slth$,
\eqn\modeldec{
\Ep(A) ~\cong~ \bigoplus_{\La\in P_+(\slth)}
( \cL(\La)\otimes \CC_{\,\La^*} )\,.}

 The space of regular polyvector fields on the base affine space, $\cP(A)$,
is shown to be a BV-algebra -- isomorphic to $\cP(\cR_3)$ -- in [\BMPbv].
It can be hoped that this reinterpretation will lead to new insights into the
geometry of $\cW$-algebras.

%
%
\appendix{A}{G-algebras, BV-algebras, and their modules}
\applab\BValgebra

\appsubsec{Preliminaries}
\appsubseclab\SSprelim

 In this appendix we will be discussing algebras $(\fA,\,\cdot)$ which
are $\ZZ$-graded, supercommutative and associative.  (Prefixes such as
super or graded will generally be implicitly understood
throughout.) Thus $\fA=\bigoplus_{n\in\ZZ}\fA^n$, with degree $|a|=n$
for $a\in\fA^n$, and the product $\cdot :\fA^m\times \fA^n \rightarrow
\fA^{m+n}$ obeys, $a\cdot b=(-1)^{|a||b|}b\cdot a$ and
$(a\cdot b)\cdot c=a\cdot (b\cdot c)$, for any homogeneous $a,b,c\in\fA$.
\smallskip

 Let us call $D: \fA^m \rightarrow \fA^{m+|D|}$ a 1st-order derivation
of degree $|D|$ if
\eqn\BVleib{
D(a\cdot b) = D(a)\cdot b + (-1)^{|a| |D|} a\cdot D(b) \, .  }  We
will refer to \BVleib\ as the Leibniz rule, and will denote by
$\cD(\fA,M)$ (or simply $\cD(\fA)$ if $M=\fA$) the space of derivations of
$(\fA,\,\cdot\,)$ with values in an $\fA$-module, $M$.
There are two distinct generalizations of the notion of a
derivation which will play a role in the following:
\item{i.} First let us define the operation of left multiplication
by $a\in\fA$, $l_a: \fA \rightarrow \fA$, via $l_a (b) = a \cdot b$.
Then \BVleib\ is clearly equivalent to the statement that for all
$a\in \fA$,
\eqn\BVnother{
D l_a - (-1)^{|a| |D|} l_a D  - l_{Da} = 0\, .
}
We may then generalize by induction, defining the map $D: \fA \rightarrow \fA$
to be an $n$th-order derivation of degree $|D|$
if $D l_a - (-1)^{|a| |D|} l_a D  - l_{Da}$ is an
$(n-1)$th-order derivation (see, \eg, [\Ko--\Ak]).
\item{ii.} A second generalization -- which we only
introduce for an $\cR$-module $M$, where $(\cR,\,\cdot\,)$ is
an Abelian algebra [\Kr] -- is the space $\cP^n(\cR,M)$ of
polyderivations of order $n$ with coefficients in $M$, defined inductively
by $\cP^0(\cR,M)\cong M$, $\cP^1(\cR,M) \cong \cD(\cR,M)$, and
$\cP^n(\cR,M)$, $n\geq 2$,  is the space of those
$a\in \cD(\cR,\cP^{n-1})$ that satisfy
$a(x,y)=-a(y,x)$, $x,y\in\cR$,
where $a(x,y)$ denotes the element $a(x)(y)\in\cP^{n-2}(\cR,M)$.

\appsubsec{Definition of G- and BV-algebras}
\appsubseclab\SSdefini

 A G-algebra [\Ge], $(\fA,\,\cdot\,,[-,-])$, is defined as a
dot algebra with the additional structure of a $\ZZ$-graded Lie
algebra under the bracket operation, $[-,-]:\fA^m\times \fA^n \rightarrow
\fA^{m+n-1}$, where
$[a,b] = -(-1)^{(|a|-1)(|b|-1)}{[b,a]}$,
such that the bracket acts as a derivation of the algebra.
It is clear from the definitions above that for any G-algebra the
subspace $\fA^0$ is an Abelian algebra with respect to the dot
product.  Similarly, $\fA^1$ is a Lie algebra with respect to the
bracket.
\smallskip

 In contrast, a BV-algebra [\BV,\Ko,\Schw] $(\fA,\,\cdot\,,\De)$
is a dot algebra with the additional structure of a
second order derivation $\De$ (BV-operator) of degree $-1$ satisfying
$\De^2=0$.
\smallskip

 There is a close relation between the two classes of algebras.  Indeed,
for any BV-algebra $(\fA,\,\cdot\,,\De)$, the bracket
\eqn\eqCc{
[a,b] ~=~ (-1)^{|a|} \left( \De(a\cdot b) - (\De a)\cdot b -
  (-1)^{|a|} a\cdot(\De b) \right)\,,\qquad a,b\in \fA\,,}
introduces on $\fA$ the structure of a G-algebra.

\appsubsec{Definition of G- and BV-modules}
\appsubseclab\Smodules

We may introduce the notion of a G- or BV-module by generalizing
the dot and bracket action of the G- and BV-algebras on themselves.
Let $\fM=\bigoplus_{n\in\ZZ} \fM^n$ be a $\ZZ$-graded module of
$(\fA,\,\cdot)$.  If $\fA$ has the additional structure of a
BV-algebra, then we define $\fM$ to be a BV-module if there further
exists a map $\De_M:\fM^n\rightarrow \fM^{n-1}$, a second order
derivation of the dot action of $\fA$ on $\fM$, such that $\De_M^2=0$.
In that case we may define a bracket operation, $[-,-]_M:\fA^m\times
\fM^n\rightarrow \fM^{m+n-1}$, by
\eqn\gmAE{
[a,m]_M ~=~ (-1)^{|a|} \big( \De_M(a\cdot m)-(\De a)\cdot m -(-1)^{|a|}
a\cdot(\De_M m)\big)\,,
}
for $a\in\fA\,,\, m\in\fM$. This bracket satisfies
\eqn\gmJI{
[a\cdot b,m]_M ~=~ a\cdot [b,m]_M+(-1)^{|a||b|}b\cdot [a,m]_M\, .
}
Moreover, the operators $[a,-]_M$, $a\in\fA$, define a representation
of the Lie algebra $(\fA,[-,-])$ and act as derivations of the
dot action of $\fA$ on $\fM$.  Any dot module $\fM$ for which a
bracket operation with these properties exists will be called a
G-module.  Clearly then, a BV-module is automatically a G-module.

\appsubsec{General examples}
\appsubseclab\SSexamples

 A standard geometric example of a dot algebra is that of smooth
polyvector fields on a given manifold, together with the wedge
product.  The Lie bracket on vector fields and functions may be
extended by induction to all polyvectors.  The resulting structure is
a G-algebra.  A straightforward abstraction of this example is the
space $\cP(\cR)=\bigoplus_{n\geq 0} \cP^n(\cR)$, of polyderivations of
an Abelian algebra $\cR$.  The algebra $(\cP(\cR),\,\cdot\,,[-.-]_S)$
is a G-algebra: the dot product is induced from that on
$\cR$ using\foot{We define, for simplicity of notation, $a(x) = 0$ for
$a \in\cR$.}
$(a\cdot b)(x)~=~a\cdot b(x) + (-1)^{|b|} a(x)\cdot b$;
and the bracket operation is the natural generalization of
the Schouten-Nijenhuis bracket [\Sch,\Ni] on polyvectors.
\smallskip

 We may also construct a large class of natural G-modules in this
context.  If $M$ is a module of $\cR$ on which the Lie algebra
$\cD(\cR)$ acts by derivations of the dot product action of $\cR$,
then the space of polyderivations $\cP(\cR,M)$ naturally has the
structure of a G-module of $(\cP(\cR),\,\cdot\,,[-,-]_S)$.  The dot
and bracket operations are induced rather similarly to the
construction of $(\cP(\cR),\,\cdot\,,[-,-]_S)$ above.

\appsubsec{Explicit examples}
\appsubseclab\bvpolyderivations

 The simplest example of the construction of a G-algebra of
polyderivations $\cP(\cR)$ as discussed above, is to take $\cR$ to be
a freely generated algebra.  Let $\cC_N\cong\CC\,[ x^1,\ldots, x^N]$
be a free Abelian algebra on $N$ generators.  It is straightforward
to verify that the G-algebra, $\cP(\cC_N)$, is nothing but the algebra
of polynomial polyvector fields on $\CC^N$, \ie,
\eqn\polasder{
\cP(\cC_N)~\cong~ \bigoplus_{n=0}^N \bigwedge{}^{\!n}\, \cD(\cC_N)\,.
}
More explicitly, it is a free $\ZZ_2$-graded algebra with
even generators $ x^1\,,\ldots, x^N\in\cC_N$ and odd generators
$ x_1^*\,,\ldots, x^*_N\in \cD(\cC_N)$, where
$x_i^*( x^j)=\de^j_i$, $i,j=1,\ldots,N$.
\smallskip

 The operator $\De_S = -\hbox{${\RSr\partial\over\LWrr{\partial
x^i}}{\RSr\partial\over\LWrr{\partial x^*_i}}$}$ is a second order
derivation on $\cP(\cC_N)$.  By direct calculation one finds that the
bracket induced by $\De_S$ is equal to the Schouten-Nijenhuis bracket; so
$(\cP(\cC_N),\,\cdot\,,\De_S)$ is a BV-algebra.  Note that there is a
canonical polyvector of maximal order, the ``volume element,'' $\Om =
\hbox{$1\over N\,!$} \ep^{i_1\ldots i_N} x^*_{i_1}\ldots x^*_{i_N}$,
and that $(\cP(\cC_N)\,,\,\cdot\,,\De_S)$, as a BV-algebra, is
generated by $ x^1\,,\ldots, x^N$ and $\Om$.
\smallskip

 For $N=2$ this is precisely the BV-algebra used in Section \AVir\ to
model the BV-algebra of operator cohomology.  Let us consider the
natural G-modules in this case.  We may introduce on $\cC_2$
two different ground ring ($\cC_2$) modules: $M_1$, isomorphic to the
ground ring itself, and the twisted module $M_{r}$ defined by the
``dual'' realization of the ground ring generators, $x_1\rightarrow
-{\p\over\p x_2}$ and $x_2\rightarrow -{\p\over\p x_1}$.  Equivalently,
we may identify $M_r$ as the space freely generated by
${\partial \over {\partial x^i}}$ on one generator, $\delta$, satisfying
$x^i \delta = 0$.  The ring $\cC_2$ now acts simply by multiplication.
Similarly, the space $\cP(\cC_2,M_r)$ is freely generated by
${\partial \over {\partial x^i}}$ and $x_i^*$ from $\delta$.  It is
clear from the discussion in Section \SSexamples\ that $\cP(\cC_2,M_r)$
is a G-module.  In fact, it is manifestly a BV-module, with the
BV-operator realized by $x^2 {\partial \over {\partial x^*_1}} +
x^1 {\partial \over {\partial x^*_2}}$.
\smallskip

 A more complicated example is given by the polyderivations of an
Abelian algebra which is not free, but whose generators satisfy a
single quadratic relation.
Consider the Abelian algebra $\cG_N=\cC_{2N}/\cI$, where $\cI$ is the
ideal generated by the vanishing relation
\eqn\vanishideal{
h_{ij}\, x^i\cdot  x^j ~=~ 0\,,}
and the metric $h$ has nonzero entries
$(h_{i (N+j)}) = (h_{(N+i)j}) = \de_{ij}$.
\smallskip

 A polyderivation $\Ph\in\cP^n(\cG_N)$ is completely
determined by its values on the ground ring generators.
Therefore, on expanding in the dual basis, $\Ph =\Ph^{i_1\ldots i_n}
x^*_{i_1}\ldots x^*_{i_n}$ where $\Ph^{i_1\ldots i_n}\in\cR_N$,
$\Ph$ is a polyderivation iff it preserves the vanishing
relation \vanishideal; \ie, the coefficients of its expansion satisfy
\eqn\realcond{
 x_i\cdot\Ph^{i\,i_1\ldots i_{n-1}} ~=~ 0\,,
\quad i_{1},\ldots,{i_{n-1}}=1,\ldots,2N\,.}
\smallskip

 The free algebra $\cC_{2N}$ carries a natural action of the
Lie algebra $\sotwon$ realized by the first order derivations
$\La_{ij} = x_i x^*_j- x_j x^*_i\,, i,j=1,\ldots,2N$.
Clearly, $\La_{ij}(\cI)\subset \cI$, so the action of
$\sotwon$ descends to the ground ring $\cG_N$, with the generators
$ x^i$ transforming in the vector ($2N$-dimensional) representation.
Using this $\sotwon$ action we can find an explicit
basis for the space of polyderivations, $\cP(\cG_N)$,
and a finite set of generators and relations which characterize
$\cP(\cG_N)$ as a dot algebra.
\thm\generat
\proclaim Theorem \generat\ [\BMPbv]. The graded, graded commutative
algebra $(\cP(\cG_N),\,\cdot\,)$ is  generated,
as a dot algebra, by $1$, the ground ring generators $ x^i$,
the order one derivation $C= x^i x^*_i$, and the order $n-1$
polyderivations
$P_{i_1\,,i_2\ldots i_n}= x_{[i_1}\vps x^*_{j_1}\ldots x^*_{j_{n}]}$,
$n=2,\ldots, 2N$,
satisfying the relations:
\eqn\conone{  x_i x^i ~=~ 0\,,}
\eqn\contwo{ x_{[i}P_{i_1,i_2\ldots i_n]} ~=~ 0\,,}
\eqn\conthr{ x^i P_{i\,,j_1\ldots j_n}~=~ -\hbox{$n\over n+1$}CP_{j_1,j_2
\ldots j_n}\,,}
\eqn\confour{P_{i_1,i_2\ldots i_m}P_{j_1,j_2\ldots j_n} ~=~
(-1)^{m-1}\hbox{$m+n-1\over n$} x_{[i_1} P_{i_2,i_3\ldots i_m]j_1\ldots
j_n}\,,}
\eqn\confive{
C P_{i_1,i_2\ldots i_{2N}} ~=~ 0\,.}
\smallskip

\noindent
These results allow us to demonstrate
that $\cP(\cG_N)$ is actually a BV-algebra by construction of
the BV-operator $\De_S$ on this basis, and
to explicitly calculate the homology of $\De_S$.  There is a unique
nontrivial homology state, the ``volume element''
$X = \textstyle{ 1\over (2N)!} \ep^{i_1 i_2 \dots i_{2N}}
  \, x_{i_1} x^*_{i_2} \dots x^*_{i_{2N}}$.
\smallskip

 For comparison with the operator cohomology of the $4D$ $\cWth$
string, the relevant case is $N=3$.  On the one hand, the ring $\cR_3$
decomposes under $\slth \subset \sosi$ as a direct sum of all
irreducible finite-dimensional modules of $\slth$, each module
occurring with multiplicity one -- \ie, $\cR_3$ is a model space for
$\slth$.  The generators $x^i$ of $\cR_3$ decompose as an $\slth$
triplet $x_\si$ and an anti-triplet $x^\si$ and satisfy the vanishing
relation $x_\si x^\si = 0$.  On the other hand, $\cR_3$ is an
irreducible representation of the algebra $\frak{so}_8$ which includes
the generators $x^i$ as well as the $\sosi$ generators $\La_{ij}$
[\BdFl].  The $\soei$ generators decompose under $\slth$ as ${\rm
ad}_{\soei}= {\bf 8}\oplus ({\bf 3}\oplus {\bf \overline 3})\oplus
({\bf 3}\oplus {\bf \overline 3})\oplus ({\bf 3}\oplus {\bf
\overline 3})\oplus {\bf 1}\oplus {\bf 1}$.  We find that there are three
ways to extend the $\sosi$ algebra to $\soei$ (${\rm
ad}_{\sosi}={\bf8}\oplus ({\bf 3}\oplus {\bf \overline 3})\oplus {\bf
1}$).  For each choice there are two ways to realize the ring
generators in terms of the remaining $\bf{3}$ and ${\bf \bar{3}}$.
Thus, there are in total six natural module structures on $\cR_3$, which
we label $M_w$, $w \in W(\slth)$.
\smallskip

  The isomorphism between these different $\soei$ realizations,
together with Theorem \generat, are enough to show that
for each $w\in W(\slth)$, the space of polyderivations with
coefficients in the ground ring module $M_w$,
$\cP(\cG_3,M_w)$, is a G-module of $\cP(\cR_3)$.
We conjecture that they are, in fact, BV-modules.  These
modules play an important role in the description of the
operator cohomology given in Section \Wthgen.

\bigskip

\noindent
{\it Acknowledgements}

 J.M.\ would like to thank the organizers of STRINGS '95
for the opportunity to present this work.  P.B.\ and J.M.\
acknowledge the support of the Australian Research Council, while
K.P.\ is supported in part by the U.S. Department of Energy Contract
\#DE-FG03-84ER-40168.

\footatend\immediate\closeout\rfile
\baselineskip=14pt{\bigskip\noindent {\bf  References}}%
\bigskip{\frenchspacing%
\parindent=20pt\escapechar=` \input refs.tmp\vfill\eject}\nonfrenchspacing


\vfil\eject\end